\documentclass[sigconf]{acmart}
\usepackage{graphicx}
\usepackage{wrapfig}
\usepackage{subcaption}
\usepackage{adjustbox}
\usepackage[utf8]{inputenc}
\usepackage{pifont}
\usepackage[T1]{fontenc}
\usepackage{tablefootnote}
\usepackage{xcolor}
\usepackage[inline]{enumitem}
\usepackage{siunitx}
\usepackage{listings}

\AtBeginMaketitle{%
\def\EDBTISSN{xxxx-xxxx}
\def\EDBTISBN{xxx-x-xxxxx-xxx-x}
\setcopyright{none}
\copyrightyear{\copyright\ 2027 Copyright held by the owner/author(s).
  Published on xxxxx under ISBN \EDBTISBN, series ISSN \EDBTISSN.
  Distribution of this paper is permitted under the terms of the
  Creative Commons license CC-by-nc-nd 4.0}
\acmYear{2027}
\acmDOI{}
\acmISBN{}

\acmConference[EDBT '27]{Extending Database Technology}{6-9 April 2027}{Lille (France)}
\settopmatter{printacmref=false, printccs=false, printfolios=false}

\newsavebox{\ximagebox}
\newlength{\ximageheight}
\newsavebox{\xglyphbox}
\newlength{\xglyphheight}
\newcommand{\xbox}[1]%
  {\savebox{\ximagebox}{#1}%
  \settoheight{\ximageheight}{\usebox{\ximagebox}}%
  \savebox{\xglyphbox}{\color{white}\char32}%
  \settoheight{\xglyphheight}{\usebox{\xglyphbox}}%
  \raisebox{\ximageheight}[0pt][0pt]{\raisebox{-\xglyphheight}[0pt][0pt]{%
    \makebox[0pt][l]{\usebox{\xglyphbox}}}}%
    \usebox{\ximagebox}%
    \raisebox{0pt}[0pt][0pt]{\makebox[0pt][r]{\usebox{\xglyphbox}}}}

\newsavebox{\LogoBox}
\sbox{\LogoBox}{}

\fancypagestyle{firstpagestyle}{%
  \fancyhf{}%
  \fancyfoot{}%
  \fancyhead[L,C]{}
  \fancyhead[R]{}{\raisebox{15pt}[0pt][0pt]{\xbox{\usebox{\LogoBox}}}}
  }

\hypersetup{%
  pdfcopyright={CC-by-nc-nd},
  pdfcreator={LaTeX with acmart, hyperref, and EDBT modifications},
  pdfvolumenum={XX},
  pdfissuenum={X},
  pdfisbn={\EDBTISBN},
  pdfissn={\EDBTISSN}
}

}% \AtBeginMaketitle

\newcommand{\cmark}{\ding{51}}

\begin{document}

%% EDBT rule: --> Please use ``titlecase'' in the title!

\title{Six Dimensions of Benchmarking Time-Series Databases}

%% EDBT rule: --> At least 1 (the corresponding) author needs to have a
%%                registered ORCID, ideally all authors have one
\author{Jalal Mostafa}
\orcid{0000-0003-2857-7816}
\affiliation{%
    \institution{Karlsruhe Institute of Technology}
    % \streetaddress{Hermann-von-Helmholtz-Platz 1}
    \city{Eggenstein-Leopoldshafen}
    % \state{BW}
    % \postcode{D-76344}
    \country{Germany}
}
\email{jalal.mostafa@kit.edu}
\authornote{Both authors contributed equally to the paper}

\author{Sandro Melissano}
\affiliation{%
    \institution{Karlsruhe Institute of Technology}
    % \streetaddress{Hermann-von-Helmholtz-Platz 1}
    \city{Eggenstein-Leopoldshafen}
    % \state{BW}
    % \postcode{D-76344}
    \country{Germany}
}
\email{sandromelissano@gmail.com}
\authornotemark[1]

\author{Nicholas Tan Jerome}
\orcid{0000-0001-5143-1183}
\affiliation{%
    \institution{Karlsruhe Institute of Technology}
    % \streetaddress{Hermann-von-Helmholtz-Platz 1}
    \city{Eggenstein-Leopoldshafen}
    % \state{BW}
    % \postcode{D-76344}
    \country{Germany}
}
\email{nicholas.tanjerome@kit.edu}

\author{Suren Chilingaryan}
\orcid{0000-0002-2909-6363}
\affiliation{%
    \institution{Karlsruhe Institute of Technology}
    % \streetaddress{Hermann-von-Helmholtz-Platz 1}
    \city{Eggenstein-Leopoldshafen}
    % \state{BW}
    % \postcode{D-76344}
    \country{Germany}
}
\email{suren.chilingaryan@kit.edu}

\author{Andreas Kopmann}
\orcid{0000-0002-2362-3943}
\affiliation{%
    \institution{Karlsruhe Institute of Technology}
    % \streetaddress{Hermann-von-Helmholtz-Platz 1}
    \city{Eggenstein-Leopoldshafen}
    % \state{BW}
    % \postcode{D-76344}
    \country{Germany}
}
\email{andreas.kopmann@kit.edu}


\renewcommand{\shortauthors}{Melissano et al.}%%MS: removed contents
%\renewcommand{\shorttitle}{} %%MS: added this one

\begin{abstract}
    Time-series databases (TSDBs) employ diverse storage architectures optimized for specific workload characteristics, leading to distinct performance profiles and bottlenecks that are often not apparent under conventional benchmarking approaches.
    System architects designing robust data backends must understand which storage engines are efficient for their particular pipelines and which exhibit the lowest risk of encountering future scalability constraints.
    This paper presents SciTSv2, a benchmarking framework that evaluates time-series databases across six workload dimensions: connection parallelism, batch ingestion, time-series regularity, multi-variate series, mixed workloads, and system metrics.
    We exploit SciTSv2 to systematically evaluate 4 TSDBs representing distinct storage engines: InfluxDB (Time-Structured Merge tree), TimescaleDB (based on relational databases), ClickHouse (columnar), and DataLayerTS (specialized in regular time-series).
    We show that each dimension surfaces architectural behavior that dedicated, single-axis benchmarks obscure, including regularity-dependent trade-offs, contention between concurrent reads and writes, and distinct CPU, I/O, and disk-bandwidth-bound bottlenecks.
    Paired with fine-grained system metrics, SciTSv2 gives architects a diagnostic tool for tracing performance outcomes back to their underlying architectural causes, supporting storage engine selections grounded in empirical, workload-specific evidence.
\end{abstract}


\maketitle

\section{Introduction}
The evolution of time-series data has gained momentum in big data environments like scientific instrumentation~\cite{katrinDes}, Internet of Things (IoT) and Industrial IoT~\cite{tsmbench, iotdb-bench, ts-iot}, IT infrastructure monitoring~\cite{ts-itinfra}, forecasting and financial trends~\cite{ts-finance}, etc.
As a response to this evolution, time-series databases (TSDBs) have emerged as a new type of database management systems to tackle the rising demands of time-series data.

The design of the novel TSDBs was motivated by the special characteristics of time-series data in comparison to other types of big data.
Time-series data are indexed using timestamps, continuously expand in size, are usually queried in ranges or summarized through aggregation and down-sampling, have very write-intensive requirements, and may be equidistant (or regular), i.e. the time interval between any 2 consecutive data points in the series is constant.
Based on these characteristics, the novel TSDBs adopt different approaches to design their storage backends yielding distinct performance measurements.
For example, DataLayerTS~\cite{dlts} is a new TSDB in the market that exploits the regularity property of time-series data to reduce stored data on disk and enable higher data ingestion rate.
Other databases like InfluxDB~\cite{influx} and ClickHouse~\cite{clickhouse} adopt their own variants of the Log-Structured Merge (LSM) tree data structure~\cite{lsmtree, lsm-survey}.

\begin{table*}[t]
    \centering
    \caption{SciTSv2 and State-of-The-Art Benchmarks Compared by The Six Dimensions}
    \label{tab:tooltable}
    \begin{tabular}{l|c|c|c|c|c|c|c||c}
        \toprule
        Dimension              & YCSB-TS\cite{ycbsts} & Smart\cite{smart-bench} & IoTDB\cite{iotdb-bench} & TS-Bench.\cite{ts-bench} & SciTS\cite{scits} & TSM\cite{tsmbench} & TSBS\cite{tsbs} & \textbf{SciTSv2} \\
        \midrule
        Time-Series Regularity &                      &                         &                         &                          &                   &                    &                 & \cmark{}         \\
        Multi-variate Series   & \cmark{}             &                         & \cmark{}                & \cmark{}                 &                   & \cmark{}           & \cmark{}        & \cmark{}         \\
        Mixed Workloads        & \cmark{}             & \cmark{}                & \cmark{}                &                          &                   & \cmark{}           &                 & \cmark{}         \\
        Connection Parallelism & \cmark{}             &                         & \cmark{}                & \cmark{}                 & \cmark{}          & \cmark{}           & \cmark{}        & \cmark{}         \\
        Batch Data Ingestion   &                      &                         & \cmark{}                & \cmark{}                 & \cmark{}          & \cmark{}           & \cmark{}        & \cmark{}         \\
        System Metrics         &                      &                         &                         & \cmark{}                 & \cmark{}          &                    &                 & \cmark{}         \\
        \bottomrule
    \end{tabular}
\end{table*}

Benchmarking TSDBs is thus important to understand the performance of TSDBs and what contributions their storage backends have to achieve better data ingestion rates and lower query latency.
In this paper, we present SciTSv2, a benchmark for TSDBs and an extension of its predecessor SciTS~\cite{scits}.
Like its predecessor, SciTSv2 focuses on heavy-write scenarios in addition to 5 queries from the scientific instrumentation real-life use-case representing raw data fetching, down-sampling, and or aggregation.
In addition to the original SciTS design, SciTSv2 adds more benchmarking dimensions to cover more storage backends and consequently more TSDBs.
In total, the benchmark adopts six dimensions to design benchmarking workloads:
\begin{enumerate}
    \item \textbf{Connection Parallelism} to study TSDB performance as function of number of connected database clients.
    \item \textbf{Batch Data Ingestion} to study how the TSDB behaves under different sizes of inserted data batches.
    \item \textbf{Regularity} as a novel dimension in TSDB benchmarking to study if exploiting time-series regularity (equidistance) can contribute to higher data ingestion and lower query latency compared to irregular time-series having variable time intervals between 2 consecutive data points.
    \item \textbf{Multi-Variate Series} to study how TSDB performance as function of how many values are stored in a single row, e.g. sensors that measure multidimensional parameters like an accelerometer measuring acceleration along the three.
          Another example is the database user's choice to store multiple data points in a single row in an attempt to increase ingestion and query performance and often leading to complex database schemas with tens to hundreds of columns.
    \item \textbf{Mixed Ingestion-Querying Workloads} to measure TSDB ingestion and query performance when the TSDB is under stress of workloads similar to those in real-life where querying and data ingestion take place concurrently.
    \item \textbf{System Metrics} to study the systematics of a TSDB storage backend and identify possible bottlenecks by collecting system metrics like CPU, memory, and disk usage while executing benchmarking workloads.
\end{enumerate}

Our benchmark is the product of hours of research in benchmarking and evaluating TSDBs for scientific instrumentation.
Existing TSDB benchmarks do not adopt all six benchmarking dimensions including the novel regularity dimension and require non-trivial modifications update, limiting the metrics the user can measure to understand and compare the performance of different TSDBs storage engines.

To demonstrate the strength of SciTSv2 design, we use it to benchmark 4 databases of four distinct storage backends: \textit{InfluxDB}~\cite{influx} to represent TSDBs based on the Time-structured Merge (TSM) tree data structure (a variant of LSM trees), \textit{ClickHouse}~\cite{clickhouse} to represent column-oriented OLAP-based TSDBs, \textit{TimescaleDB}~\cite{timescale} to represent TSDBs based on traditional relational DBMSs, and DataLayerTS~\cite{dlts} to represent storage backends specializing in regular time series.
The paper's contributions can be summarized as:
\begin{enumerate}
    \item We present the design and the implementation of SciTSv2 as a benchmark for TSDBs that supports the discussed six benchmarking dimensions for heavy-write operations in addition to 5 real-life practical queries from the scientific instrumentation use-case~\cite{scits}.
    \item A deep performance evaluation based on the six benchmarking dimensions is performed on the 4 selected TSDBs.
          We provide insights on their performance and their underlying indexing and storage techniques.
    \item The impact of large multi-variate series on the performance of the selected databases is studied to show the trade-offs between the database performance and the complexity of its schema.
    \item Exploiting the regularity property in time-series data appears as a very promising approach for use cases with equidistant time series.
          No such storage engines have been benchmarked yet to our knowledge.
          We benchmark DataLayerTS, an emerging array-based TSDB for equidistant time series and provide insights on the performance behavior of every architecture of the 4 selected databases under regular and irregular time-series workloads.
    \item We perform quantitative and qualitative analysis of system resource consumption and performance metrics of the 4 different selected TSDBs to study how storage backends run into bottlenecks with increasing requirements.
\end{enumerate}

The rest of this paper is organized as follows:
Section~\ref{sec:related-works} shows the related works and the state of the art research in TSDB benchmarking.
Section~\ref{sec:background} explains some background material about the use-case of SciTSv2 and the architecture of the benchmarked TSDBs.
Section~\ref{sec:arch} illustrates the architecture and the design of SciTSv2 showing its strengths and limitations.
Section~\ref{sec:experimental-setup} shows the experimental setup we used to show the strengths of SciTSv2 desgin.
Similarly, Section~\ref{sec:results} shows the results of the 4 benchmarked TSDBs.
We finally conclude in Section~\ref{sec:conclusion}.

\section{Related Works}
\label{sec:related-works}

Several benchmarking frameworks have been proposed for time-series databases, yet none systematically covers all six dimensions we identify.
\autoref{tab:tooltable} summarizes the coverage of existing benchmarks against our dimensions: time-series regularity, multi-variate series, mixed workloads, connection parallelism, batch data ingestion, and system metrics.

YCSB-TS extends the YCSB framework to time-series workloads but focuses primarily on basic read/write operations, omitting regularity, mixed workloads, batch ingestion, and system metrics~\cite{ycbsts}.
TSBS supports realistic IoT and DevOps scenarios and includes connection parallelism, multi-variate time-series, and batch ingestion, but does not consider regularity or mixed workloads, nor does it collect system-level resource metrics~\cite{tsbs}.
TS-Bench provides systematic query and ingestion testing for monitoring wind turbines but similarly lacks regularity and mixed-load dimensions~\cite{ts-bench}.
IoTDB-Benchmark targets IoT deployments with high-velocity data, yet its scope excludes time-series regularity and collection of system metrics~\cite{iotdb-bench}.
While IoTDB-Benchmark supports quantifying the performance for equidistant time-series arriving out of its initial order, it does not support non-equidistant irregular time-series.
SmartBench focuses on spatial-temporal data in smart environments, covering mixed workloads but not regularity, multi-variate time-series, or detailed system metrics~\cite{smart-bench}.
TSM-Bench addresses monitoring applications with emphasis on queries in the context of hydrology, mixed workloads and connection parallelism, but does not exploit regularity or system metrics~\cite{tsmbench}.
The original SciTS~\cite{scits} introduced connection parallelism, batch ingestion, and system metrics for scientific instrumentation, but did not support regularity, multivariate data, or mixed ingestion-querying.

As shown in \autoref{tab:tooltable}, no prior benchmark combines all six dimensions.
In particular, time-series regularity -- critical for array-based stores like DataLayerTS -- has not been investigated before.
SciTSv2 is the first to unify all six dimensions and enable quantifying performance of TSDBs under time-series regularity workloads, enabling direct comparison of storage architectures under diverse, realistic conditions that reflect the complexity of scientific monitoring and industrial IoT systems.

\section{Background}
\label{sec:background}
This section explains some background material about the scientific instrumentation use-case which we derive the benchmark queries from and gives relevant information about the architecture of storage backends for the 4 benchmarked TSDBs.

\subsection{Scientific Instrumentation Use-case}
\label{sec:usecase}
Scientific experiments produce and store large amounts of time-series data that is relevant to the operation and maintenance of the operating instrumentations.
For example, it can incorporate operational and environmental measurements contributing to a scientific observation or to detecting systems failures.
All of this data is collected and archived in permanent storage by the experiment's monitoring subsystem for later analysis.

These instrumentations can be composed of thousands of sensors producing tens of time-series measurements per second for each sensor and active all over the year.
Such a use-case raise the concerns regarding the storage and the retrieval of continuously expanding time-series data~\cite{katrinDes}.
For example, the KArlsruhe TRItium Neutrino (KATRIN) experiment uses relational Microsoft SQL and MySQL databases to store and query time-series data from around 100,000 sensors with \SI{10}{\hertz} to \SI{0.1}{\hertz} sampling frequencies~\cite{katrinDes}.
With the growing data volumes and rates at KATRIN, the ACID constraints of the relational databases introduced a performance bottleneck when scaling the data ingestion and retrieval workloads~\cite{nosqleval_cmu,traditionaldbms,mongodb_sensors}.
Therefore, scientists at KATRIN are reviewing TSDB as a promising alternative for the operating RDBMSs.
SciTS was initially designed to help scientists choose a TSDB for KATRIN's use-case, its queries were inspired by KATRIN workloads especially.
SciTSv2 build on top of the initial design to add three more benchmarking dimensions: Mixed Workloads, Regularity, and Multi-Variate Time-Series.

\subsection{Architecture of TSDBs' Storage Engines}
\label{sec:tsdb-architectures}
TSDBs storage engines have followed different approaches to design high-performance scalable databases capable of satisfying time-series data requirements.
This paper selects 4 TSDBs of four distinct storage engines for performance evaluation:

\paragraph{TimescaleDB (TS)}
It is an extension for the relational row-based PostgreSQL.
It exploits PostgreSQL B-Tree-based indexing and SQL queries.
However, it partitions one PostgreSQL table into smaller chunks called hypertables based on the time-series data timestamp, e.g. a chunk for every 7 days.
Indexing and writing to chunks that can fit in the main memory boost ingestion rates and query latency in comparison to traditional PostgreSQL design.
For lower query latency and lower storage overhead, TimescaleDB deploys a built-in background job to perform asynchronous age-based data compression by transforming the hypertables rows (hot data) into a compressed columnar format (cold data).

\paragraph{InfluxDB (IF)}
The storage engine of Influx v2 is based on the \textit{Time-Structured Merge (TSM)} tree data structure, a LSM variant specialized in time-series data~\cite{lsmtree}.
Inserted data in TSM trees is written to a Write-Ahead Log (WAL) first and then copied to the cache memory while maintaining indexes in memory.
The time-series data is persisted on the storage using immutable shards, each shard contains the data of a corresponding duration of time.
An InfluxDB data point consists of a timestamp, a value, and one or more tags.
Tags are key-value pairs that are used to add more information to the data point.
InfluxDB uses timestamps and tags for indexing.
It uses per-data-type compression algorithms e.g. ZigZag encoding for integers, the Gorilla algorithm~\cite{gorilla} for float numbers, simple8b~\cite{simple8b} for timestamp indexes, bit packing for booleans, and the snappy algorithm~\cite{snappy} for strings.

\paragraph{ClickHouse (CH)}
A column-based store OLAP DBMS that is designed for high ingestion rates using a variant of LSM trees called \emph{MergeTree}.
The MergeTree data structure writes the time-series data directly to permanent storage as multiple partitions to enable unrestricted data ingestion.
CH organizes access to these partitions when queried until a background job sorts and merges all partitions into one.
This enables CH to use efficient sparse indexing to locate data in the partitions quickly.
While CH was designed primarily for big data~\cite{clickhouse_bigdata}, CH has very good support for time-series data enabling low-latency queries through its built-in time-series processing functions.

\paragraph{DataLayerTS (DLTS)}
A vector-store specialized in regular or equidistant time-series.
Instead of saving a timestamp for every data point in the time-series, DLTS saves the timestamp of the first data point of the sensor and the time resolution of the series.
This enables higher ingestion rates and lower query latency because the storage engine can write and read less information to and from the disk.
Every sensor has a private WAL where DLTS writes the data to first.
Data from the private WAL is compressed using the Brotli data format~\cite{brotli} and then added to a \texttt{double float} array for each sensor by a background process.
Additional data other than values is treated as metadata and is stored seperately.
Querying a specific time-value pair involves unpacking the whole value array, extracting the corresponding metadata, and calculating the timestamp using the data index in the values array and the first recorded timestamp of the series in addition to the configured time resolution.
DLTS does support irregular time-series by storing a timestamp for each data point in the series through a specialized Application Programming Interface (API), but DLTS developers do not recommend it for intense workloads due to its performance impact.

\section{The SciTSv2 Benchmark}
\label{sec:arch}

\begin{table*}[!ht]
    \caption{User-defined Parameters of SciTSv2 Workloads}
    \label{tab:parameters}
    \begin{tabular}{lp{9cm}c}
        \toprule
        Name                     & Description                                                           & Workload Type    \\
        \midrule
        TargetDatabase           & The type of the target database server e.g. InfluxDB, ClickHouse, etc & Ingestion/Query  \\
        DaySpan                  & Length of the whole time-series in the database table in days         & Ingestion/Query  \\
        StartTime                & Earliest timestamp to be stored into or retrieved from the database   & Ingestion/Query  \\
        BatchSizeOptions         & Size of batch to insert into table                                    & Ingestion        \\
        ClientNumberOptions      & Number of concurrent clients                                          & Ingestion        \\
        SensorNumber             & Number of sensors to simulate to represent cardinality                & Ingestion        \\
        QueryType                & An enum representing the query type e.g. Q1-Q5                        & Query            \\
        TestRetries              & How many times to repeat the query test                               & Query            \\
        DurationMinutes          & Length of time-series data in minutes                                 & Query (Q1 to Q5) \\
        AggregationIntervalHour  & Length of time window to apply the down-sampling function on          & Query (Q3 to Q5) \\
        SensorsFilter            & A list of sensor IDs to filter on in the query                        & Query (Q1 to Q5) \\
        MaxValue                 & The upper boundary of the sensor's value used in Q2                   & Query            \\
        MinValue                 & The lower boundary of the sensor's value used in Q2                   & Query            \\
        IngestionType            & Regularity in data generation for time-series ingestion               & Ingestion        \\
        MixedWLPercentageOptions & Ingestion-to-query percentage of a mixed workload                     & Ingestion/Query  \\
        DataDimensionsNrOptions  & Number of variables in a multi-variate series                         & Ingestion/Query  \\
        Mode                     & Benchmarking Mode whether ingestion, query, or mixed workloads        &                  \\
        \bottomrule
    \end{tabular}
\end{table*}

The initial version of SciTS~\cite{scits} focused on high-performance ingestion and query workloads of irregular time-series data.
SciTS provides multi-threaded asynchronous workload execution including batch-based data ingestion to stress the target TSDBs with parallel connections and higher data volumes.
Sytem metrics are first-class citizens in SciTS by implementing a Glances client that collects the metrics from a Glances server~\cite{glances} running on the target host.
Thus, SciTS supported 3 out of the 6 dimensions: connection parallelism, batch data ingestion, and system metrics.

\begin{figure}[ht!]
    \includegraphics[width=\linewidth]{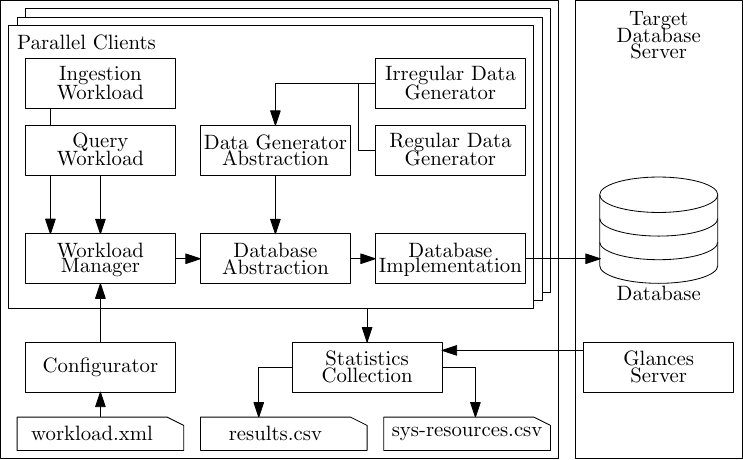}
    \caption{SciTSv2 Architecture}
    \label{fig:scitsv2-arch}
    \Description{}
\end{figure}

With regular and irregular time series are both available in real-life scenarios~\cite{tsbs,iotdb-bench} and the rise of new databases that exploit time-series regularity in its storage engine, benchmarking TSDBs while considering regularity has become a necessity.
Similarly, mixed ingestion and query workloads in addition to multi-variate time-series are foundational to real-life workloads.
Therefore, a good TSDB benchmark should support the other 3 dimensions: mixed workloads, time-series regularity, and multivariate time-series.
For this purpose, we exploit the highly extensible and customizable architecture of SciTS to bring SciTSv2 to life.

\autoref{fig:scitsv2-arch} shows the architecture of SciTSv2.
The configurator reads the workload definition file and launches at least one database client to start the benchmarking process.
The workload manager of each client configures the ingestion and queries workload components to generate the final workload simulating any specified benchmarking requirement.
SciTSv2 supports mixed workloads by mixing components of the ingestion and the query workloads in the workload manager.
Hereby, we call pure ingestion or pure query workloads ``Dedicated Workloads'', and a combination of concurrent ingestion and query ``Mixed Workloads''.
All interactions with the TSDB are managed by the database abstraction layer that enables SciTSv2 to treat all TSDBs alike.
Each TSDB defines its implementation of the database abstraction layer based on its data manipulation language (DML).
Workloads that has an ingestion component use SciTSv2's random time-series data generators that simulate sensors in the scientific instrumentation use-case.
SciTSv2 has 2 multivariate time-series data generators: regular and irregular time-series data generators.
On the other hand, the irregular time-series generator creates a time-series of non-equidistant incremental time-stamps.
Using the data generator abstraction layer is unified through the data generator abstraction layer.
The benchmark collects system metrics like CPU, memory, disk, and network usage by collecting statistics from a Glances server running on the target database host.
Collected system and benchmarking metrics are saved in 2 separate CSV files at the end of execution.

The definition of mixed workloads can be done in different ways.
For example, they can be distributed to some ratio of ingestion clients and query clients such as 50\% of parallel clients are performing ingestion and 50\% are executing queries.
Instead, SciTSv2 distributes ingestion and query workloads in mixed workloads based on the number of inserted or queried data points.
This desgin bypasses possible bottlenecks in TSDBs while handling parallel clients.
It fairly measures the reaction of a TSDB in mixed workloads by focusing on the real goal of a TSDB: to manage time-series data points.

% In the world of time-series data, data points in a time-series are not guaranteed to arrive in order.
% The reasons can be numerous: distributing communications between the TSDB client and server on multiple parallel TCP connections, or distributed message brokers (e.g. Apache Kafka) which can re-arrange the data flow between the TSDB client and server, etc.
% Out-of-order data can impact the time-series regularity from the TSDB server point of view, and thus impacting the TSDB performance.
% SciTSv2 simulates such cases by supporitng 2 types of time-series data flows in the regular and irregular time-series data generators:
% \begin{enumerate*}
%     \item \textbf{Patchwork}: out-of-order batches of regular or irregular time-series data to simulate batches arriving out-of-order to the TSDB (data inside one batch is still considered ordered),
%     \item \textbf{Contiguous}: ordered batches of regular or irregular time-series data where time-series data is sent perfectly ordered inside a batch and among all batches.
% \end{enumerate*}

\subsection{Supported Queries}
\label{sec:queries}
Inspired by the {KATRIN}'s scientific instrumentation, SciTSv2 extends the proposed queries in SciTS~\cite{scits} to support multi-variate time-series in some queries.
The queries represent raw, aggregated, or down-sampled data of one or more sensors.
We assume the following table schema in our queries: \textit{(time\_field,~sensor\_id,~value\_field\_0,~\dots,~value\_field\_N)}.
% The function \textit{TRUNCATE} in our queries returns a list of time intervals of a specified length, e.g. \textit{TRUNCATE(`1min', time\_field)} returns a list of time-intervals where each item represents a 1-minute of data using the column \textit{time\_field}.
The queries and their SQL equivalents can be described as follows:

\begin{enumerate}%[label=\textbf{(Q\arabic*)}]
    \item[\textbf{(Q1-A)}] Mono-Variate Raw Data Fetching: Get the raw value of the first dimension of one or more sensors over a duration of time. It is used to visualize and analyze raw data of specific sensors with non-dimensional data.

          %           \begin{lstlisting}
          % SELECT dimension_1 FROM table 
          % WHERE time_field > ? AND time_field < ? 
          %     AND sensor_id = ANY(?, ?, ?, ...)
          % \end{lstlisting}
    \item[\textbf{(Q1-B)}] Multi-Variate Raw Data Fetching: Get the raw values of all dimensions of one or more sensors over a duration of time. It is used to visualize and analyze raw data of multidimensional sensors.

          %           \begin{lstlisting}
          % SELECT *  FROM table
          % WHERE time_field > ? AND time_field < ?
          %     AND sensor_id = ANY(?, ?, ?, ...)
          % \end{lstlisting}
    \item[\textbf{(Q2)}] Out of Range Query: Get the intervals over a duration of time where the value of a specific sensor was out of a defined range. It is used to detect when the sensor was acting abnormally in a specific interval of time. The value filtering is done using a single variable of the data point in case of multi-variate time-series.

          %           \begin{lstlisting}
          % SELECT TRUNCATE(period, time_field) AS interval, 
          % MAX(value_field), MIN(value_field) FROM table
          % WHERE time_field >= ? AND time_field <= ?
          %     AND sensor_id = ? 
          % GROUP BY interval HAVING MIN(value_field) < ?
          % OR MAX(value_field) > ?
          % \end{lstlisting}

    \item[\textbf{(Q3)}] Data Aggregation: Represent the data of one or more sensors over a specific duration of time using one aggregated value of an aggregation function denoted by \textit{agg\_func} e.g. the standard deviation, the mean, etc. Aggregation is performed using a single variable of the data point in case of multi-variate time-series.

          %           \begin{lstlisting}
          % SELECT agg_func(value_field) FROM table
          % WHERE time_field >= ? AND time_field <= ?
          %     AND sensor_id = ANY(?, ?, ?, ...)
          % \end{lstlisting}

    \item[\textbf{(Q4)}] Data Down-Sampling: down-sample one or more sensors using a specific sampling function denoted by \textit{agg\_func} over a duration of time. Aggregation is performed using a single variable of the data point in case of multi-variate time-series.

          %           \begin{lstlisting}
          % SELECT TRUNCATE(period, time_field)
          % AS interval, sensor_id, agg_func(value_field)
          % FROM table WHERE time_field >= ? AND time_field <= ?
          %     AND sensor_id = ANY(?, ?, ?, ...)
          % GROUP BY interval, sensor_id
          % \end{lstlisting}

    \item[\textbf{(Q5)}] Operations on Two Down-sampled Sensors: Downsample the data of two sensors over a duration of time and using the function \textit{agg\_func}, then compare the results using the function \textit{comp\_func}. A use case of this query is comparing the data of two down-sampled sensors using value subtraction. Aggregation is performed using a single variable of the data point in case of multi-variate time-series.

          %           \begin{lstlisting}
          % SELECT Sensor1.period, comp_func(Sensor1.val, Sensor2.val)
          % FROM
          %   (SELECT TRUNCATE(period, time_field)
          %    AS interval, agg_func(value_field) AS val
          %    FROM table WHERE time_field >= ?
          %     AND time_field <= ?
          %     AND sensor_id = ANY(?, ?, ?, ...)
          %    GROUP BY interval)Sensor1
          % INNER JOIN 
          %   (SELECT TRUNCATE(period, time_field)
          %    AS interval, agg_func(value_field) AS val
          %    FROM table WHERE time_field >= ?
          %     AND time_field <= ?
          %     AND sensor_id = ANY(?, ?, ?, ...)
          %    GROUP BY interval)Sensor2
          % ON Sensor1.period = Sensor2.period
          % \end{lstlisting}
\end{enumerate}
\subsection{Workload Definition}
A workload is a set of benchmarking parameters and the connection details of the target database server in a XML configuration file as shown in \autoref{tab:parameters}.
An ingestion workload is defined by parameterizing SciTSv2 using:
\begin{enumerate*}[label=(\arabic*)]
    \item \textit{ClientNumberOptions} to represent concurrency i.e. the number of database clients to insert records into the database;
    \item \textit{BatchSizeOptions} to configure the batch size to insert in one operation,
    \item \textit{SensorNumber} to parameterize the cardinality of the database table by configuring a specific number of sensors;
    \item \textit{DataDimensionsNrOptions} to parametrize the cardinality of the single data point, transforming it into a multivariate data point;
    \item \textit{IngestionType} to set the regularity options of the time-series;
    \item and \textit{MixedWLPercentageOptions} defines the intensity of concurrent retrieval workloads to simulate specific I/O scenarios.
\end{enumerate*}
For instance, a \emph{connection parallelism} workload is defined by setting the \textit{ClientNumberOptions} to a set of the total parallel connections to test with, e.g. setting it to 1,2,4 will run the same workload with 1 database client, then 2 parallel clients, then 4 parallel clients in a single benchmark run.
Similarly, the \emph{batch data ingestion} workload sets \textit{BatchSizeOptions} to a set of batch sizes in a single benchmark run.
The user can simulate any ingestion workload by making the desired paramters combination.

For mixed or query workloads, the user can use the configuration file to set the target query using the \textit{QueryType} option.
The five queries can be parameterized by choosing the queried time intervals (\textit{DurationMinutes} in \autoref{tab:parameters}), and by filtering on one or more sensors using the \textit{SensorsFilter} parameter.
Down-sampling and aggregation queries are additionally parameterized by specifying an aggregation or a sampling interval.
The benchmark uses the \textit{average} function to calculate aggregations in the aggregated or sampled time interval.
Other queries like out-of-range queries that require filtering on the \textit{value} columns can be parameterized in the configuration file using the \textit{MinValue} and \textit{MaxValue} fields.
To assess the results correctness, the user can repeat the same query with the same parameters as much as needed using the \textit{TestRetries} parameter.

The regularity of generated time-series is specified using the parameter \textit{IngestionType}.
Mixed workloads can be performed by a specifying the percentage of ingestion-to-query data points using the \textit{MixedWLPercentageOptions} parameter.
The parameter \textit{DataDimensionsNrOptions} defines the number of variables in a multi-variate series used for both ingestion and queries.

\subsection{Implementation}
SciTSv2 has an open-source implementation\footnote{https://github.com/sandrosano/scits} in cross-platform C\#.
Its implementation is highly extensible due to its object-oriented design, abstraction layers, and resilient configuration.
The benchmark can support any TSDB as long as the user implements the database abstraction layer, as shown in \autoref{fig:scitsv2-arch}, using the TSDB DML.
For best performance, SciTSv2 is highly parallel and uses non-blocking asynchronous I/O operations for interactions with the target database server.

Its random data generator generates timestamps incrementally based on the date and periods defined in the workload definition file.
Sensors' values are considered to be random values that are uniformly ranging between zero and the max value of a signed 32 bits integer.

\section{Experimental Setup}
\label{sec:experimental-setup}
To show the strengths of SciTSv2, we use it to benchmark the following TSDBs: InfluxDB, TimescaleDB, ClickHouse, and DataLayerTS.
Our experimental setup is the first that considers a TSDB specialized for regular time-series workloads, namely DataLayerTS.

We run our benchmarks in IONOS public cloud where 2 dedicated server-class machines are rented and exclusively used for our experiments.
To run the target TSDB server, the \emph{server} machine uses an Intel(R) Xeon(R) Platinum 8370C CPU @ 2.80GHz with 8 logical cores and \SI{48}{\mega\byte} L3 cache, 32 GB DDR4 ECC, and 1 TB NVMe SSDs formatted with XFS filesystem.
To record system metrics, we also run a Glances server on the \emph{server} machine.
SciTSv2 runs on the \emph{client} machine which uses an Intel(R) Xeon(R) Platinum 8370C CPU @ 2.80GHz with 64 logical cores and 2 instances of \SI{96}{\mega\byte} L3 cache distributed over 2 NUMA nodes, 512 GB DDR4 ECC.

We configure the ClickHouse server to partition data every day. Each partition is then ordered by the table's primary key the tuple \textit{(timestamp, sensor\_id)}.
Indices are defined on both of the fields: \textit{timestamp}, and \textit{sensor\_id}.
We use ClickHouse v22.1.3.7 with its native TCP protocol and we set the following configurations: \textit{max\_server\_memory\_usage\_to\_ram\_ratio} to 0.9, \textit{async\_insert} is off, and \textit{index\_granularity} is 8192 rows.

We use InfluxDB v2.1.1 and the \textit{Line} protocol to insert data and the server is set up with the following configuration: \textit{storage-wal-fsync-delay} is set to 0, \textit{storage-cache-max-memory-size} is set to \SI{1048}{MB}, and \textit{storage-cache-snapshot-memory-size} is set to \SI{100}{MB}.

Based on TigerData's recommendations, a TimescaleDB v2.5.1 server is configured with a \textit{hypertable} of a 12-hours chunking interval so chunks constitute no more than 25\% of the main memory.
TimescaleDB compression is configured to compress row data (or hot data) into the columnar format (or cold data) every 7 days of data and to order the columnar data by \textit{timestamp} and \textit{sensor\_id}.
The server is configured with the pgtune-based tool \textit{timescale-tune}, we set the following parameters: \textit{shared\_buffers} to \SI{7994}{MB}, \textit{maintenance\_work\_mem} to \SI{2047}{MB}, and \textit{max\_parallel\_workers} to 8 workers, and we use PgBouncer as a connection pooler to manage parallel connections.

DataLayerTS is designed for regular time-series.
While it supports irregular time-series for compatibility reasons, it is not designed for such workloads.
Benchmarking DLTS with irregular time-series will lead to very low ingestion rate and high query latency.
For a fair comparison, we do not consider such evaluations.

\section{Experimental Evaluation}
\label{sec:results}
This section uses SciTSv2 design to perform a performance evaluation and analysis of the 4 targeted TSDBs.
We perform analysis of dedicated and mixed workloads in addition to insights from system metrics like CPU, memory, and disk usages.

\subsection{Dedicated Ingestion Workloads}
\label{sec:dedicated-ing}

\subsubsection{Mono-Variate Regular and Irregular Time Series}
\label{sec:monovariate}
We measure the ingestion rate of a dedicated ingestion workload where no queries are executed concurrently of a mono-variate time-series.
The measured rate is studied as function of the inserted batch size, number of parallel clients performing ingestion, and time-series regularity.

\begin{figure}[!ht]
    \centering
    \includegraphics[width=\linewidth]{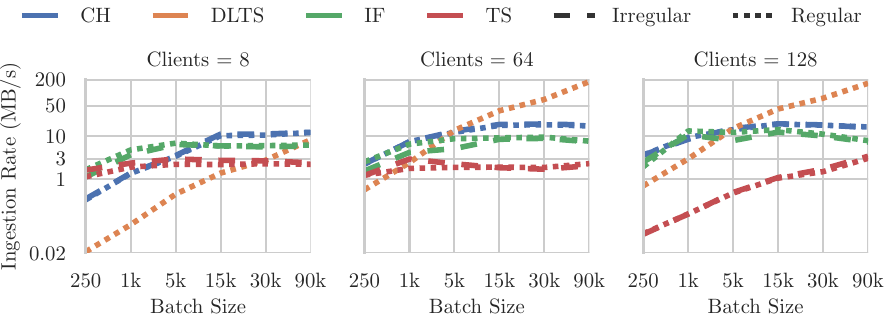}
    \caption{Ingestion Rate of Regular and Irregular Mono-Variate Time-series as function of Batch Size and Parallel Clients}
    \label{fig:reg-bscn}
    \Description{}
\end{figure}

In \autoref{fig:reg-bscn}, we see that CH, DLTS, and IF achieves higher ingestion rate.
TS ingestion rate ranges between \SI[per-mode=symbol]{1}{\mega\byte\per\second} at small batch sizes and \SI[per-mode=symbol]{4}{\mega\byte\per\second} at high batch sizes.
However, as we increase the number of parallel clients, its ingestion rate may drop below $\sim$\SI[per-mode=symbol]{1}{\mega\byte\per\second} which may be caused by deficiencies in PostgreSQL's design managing concurrent connections despite using the PgBouncer connection proxy.
The ingestion rate of IF ranges between \SI[per-mode=symbol]{1}{\mega\byte\per\second} at small batch sizes and \SI[per-mode=symbol]{10}{\mega\byte\per\second} at high batch sizes.
Other TSDBs like DLTS and CH have better ingestion rate with the same CPU resources to scale the ingestion rate as the number of parallel connections increase.
CH achieves a rate of up to $\sim$\SI[per-mode=symbol]{21}{\mega\byte\per\second} due to the low overhead of its columnar storage engine and its sparse indexing algorithms.
As we increase the client number to 64 and 128, DLTS can reach the rate of $\sim$\SI[per-mode=symbol]{178}{\mega\byte\per\second} with large batches.
DLTS takes advantage of the time-series regularity where its storage engine writes less bytes per data point to the disk by eliminating the need to write and index the data points' timestamps.
As seen in \autoref{fig:reg-bscn}, DLTS ingestion rate is much higher on larger batch sizes than small batch sizes.
Under small number of parallel data clients or small batch sizes, DLTS's ingestion rate drops even below \SI[per-mode=symbol]{1}{\mega\byte\per\second}.
In these circumstances, DLTS storage engine performs more checkpoints to merge time-series data in the WAL with the final historical array.
The merging process re-writes the full array on every checkpoint.
An increasing number of checkpoints can consequently drop the ingestion rate of DLTS.

% DLTS can be configured to decide when a checkpoint is executed, however, that would result in higher 
% This saves write-amplification (I/O-Usage), as for every checkpoint, the array needs to be fully rewritten, which is the weak point of DLTS. Rising the checkpointer frequency will improve speed for small batch size configurations, but results in a tinier I/O-bottleneck and lowers DLTS's speed for big batches. As we see in Figure \ref{fig:sdisk}, write amplification is an important weakness in DLTS.
CH, IF, and TS do not show a clear pattern when comparing ingestion rate of regular and irregular time-series.
We also observe that these TSDBs reach the upper limit of their ingestion rate around the batch size of 15000.
Only DLTS can scale further when using higher batch sizes.

\begin{figure}[!ht]
    \centering
    \begin{subfigure}[b]{0.49\linewidth}
        \centering
        \includegraphics[width=\linewidth]{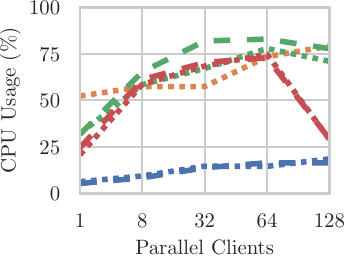}
        \caption{CPU and Parallel Clients}
        \label{fig:cpu-cn}
    \end{subfigure}
    \begin{subfigure}[b]{0.49\linewidth}
        \centering
        \includegraphics[width=\linewidth]{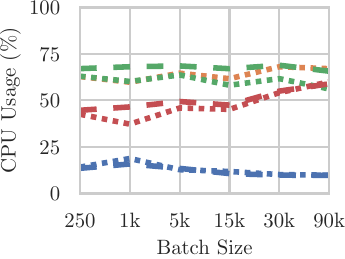}
        \caption{CPU and Batch Sizes}
        \label{fig:cpu-bs}
    \end{subfigure}
    \hfill
    \begin{subfigure}[b]{0.49\linewidth}
        \centering
        \includegraphics[width=1\linewidth]{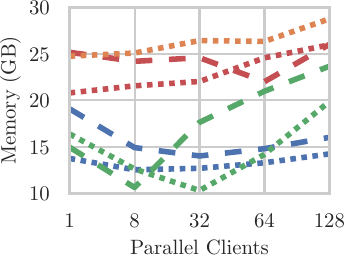}
        \caption{Memory and Parallel Clients}
        \label{fig:mem-bs}
    \end{subfigure}
    \begin{subfigure}[b]{0.49\linewidth}
        \centering
        \includegraphics[width=1\linewidth]{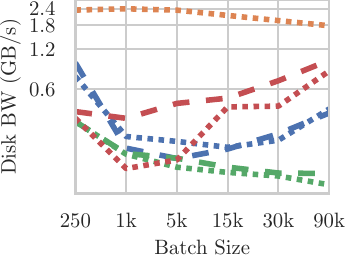}
        \caption{128 Clients Disk Bandwidth}
        \label{fig:sdisk}
    \end{subfigure}
    \caption{System Resources for Regular and Irregular Time-series Ingestion.}
    \label{fig:sys-monoingestion}
    \Description{}
\end{figure}

To understand the ingestion bottlenecks of each TSDB, we look at their consumed system resources in regular and irregular data ingestion, as shown in \autoref{fig:sys-monoingestion}.
\autoref{fig:cpu-cn}, \autoref{fig:cpu-bs} and \autoref{fig:mem-bs} shows the CPU and memory usage as function of parallel clients and batch size.
IF, TS, and DLTS are constantly high on CPU usage reaching more than 75\% at some points.
However, the maximum CPU usage of CH is still below 25\% for the same workloads.
This is also true when comparing memory of CH with TS, IF, and DLTS.
Considering CH's competitive ingestion rate, it is remarkably resource-efficient.
This is because of its MergeTree storage architecture which prioritizes writing data directly to disk and because its lightweight sparse indexing algorithm that does not consume CPU resources to compute time-series indices.
In most setups, IF have the highest CPU usage among the 4 TSDBs to sort the values before writing them to persistent storage as TSM files.

Disk I/O is usually considered one of the most critical limitations in scalable high-available systems like TSDBs.
To understand how efficient the target TSDBs use disk I/O, we analyze the disk writing bandwidth.
Figure \ref{fig:sdisk} shows the average disk write bandwidth for each TSDB, measured in megabytes per second.
We notice that DLTS's disk write bandwidth ranges between \SI[per-mode=symbol]{2}{\giga\byte\per\second} and \SI[per-mode=symbol]{2.5}{\giga\byte\per\second} even at small batch sizes.
Other TSDBs like CH and IF consume less than \SI[per-mode=symbol]{500}{\mega\byte\per\second}.
Contrarily, TS has a higher disk BW than CH and IF (up to \SI[per-mode=symbol]{1}{\giga\byte\per\second}) and less ingestion rate (up to \SI[per-mode=symbol]{4}{\mega\byte\per\second}) than both of them as shown in \autoref{fig:reg-bscn}.
TS's disk usage is considered high in compariso to its low ingestion rate.
The high disk usage is driven by managing and writing a dedicated B-Tree index for each chunk of the TS's hypertable.

\begin{figure}[!ht]
    \includegraphics[width=\linewidth]{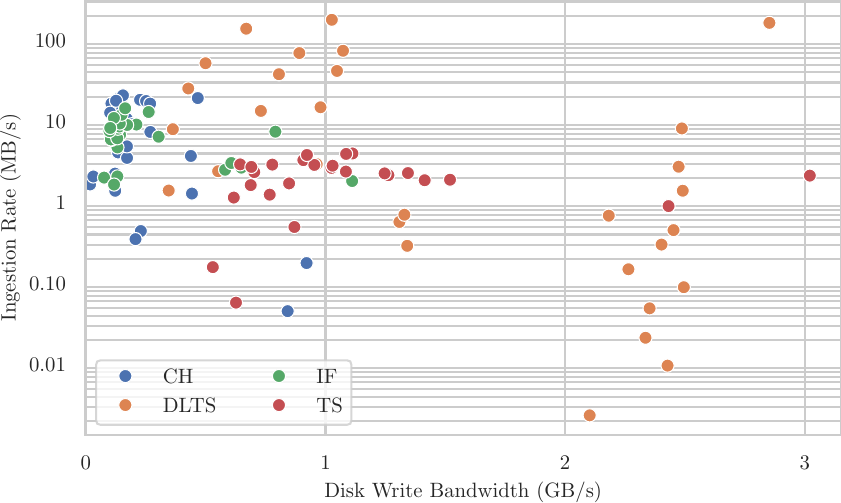}
    \caption{Ingestion Rate against Disk Write Bandwidth for All Mono-Variate Dedicated Ingestion Experiments with Different Parallel Clients and Batch Size Combinations}
    \label{fig:bs15-Disk-bandwith+NW-vs-Bandwidth-a}
    \Description{}
\end{figure}

Some TSDBs, like DLTS and TS, consumes higher disk write bandwidth than others, as shown in \autoref{fig:sdisk}.
To understand the reasons, we plot the ingestion rate as function of disk write bandwidth for all mono-variate dedicated ingestion experiments with different combinations of parallel clients~(1, 8, 32, 64, 128) and batch sizes~(250, 1k, 5k, 15k, 30k, 90k), as shown in \autoref{fig:bs15-Disk-bandwith+NW-vs-Bandwidth-a}.
Both IF and CH shows the best disk usage efficiency when we compare their ingestion rate with disk write bandwidth.
On the contrary, DLTS and TS shows that their disk write bandwidth is higher than IF and CH without a clear correlation with their ingestion rate.
This phenomenon is known as Disk Write Amplification where the actually written data size is multiple of the logical size intended to be written.
For example, at DLTS's maximum rate $\sim$\SI[per-mode=symbol]{178}{\mega\byte\per\second}, the actual written data size was \SI[per-mode=symbol]{1}{\giga\byte\per\second}, $5.7\times$ more than data intended to be written.
This is caused by how DLTS merges WAL data with the historical array.
On every checkpoint, DLTS re-writes the full historical array with the new appended values.
This inefficient merging algorithm leads to high disk usage as a result, and consequently Disk Write Amplification.
Similarly, TS's data ingestion rate is nearly constant at $\sim$\SI[per-mode=symbol]{4}{\mega\byte\per\second} while its disk write bandwidth ranges between $\sim$\SI[per-mode=symbol]{0.5}{\giga\byte\per\second} and $\sim$\SI[per-mode=symbol]{3}{\giga\byte\per\second}.
TS writes the time-series as a chunked hypertable, each chunk of the hypertable is a PostgreSQL table with its own B-Tree index.
Thus, increasing the amount of data written to the disk, and consequently leading to Disk Write Amplification.

\subsubsection{Multi-Variate Time-Series}
\begin{figure}[!ht]
    \centering
    \includegraphics[width=1\linewidth]{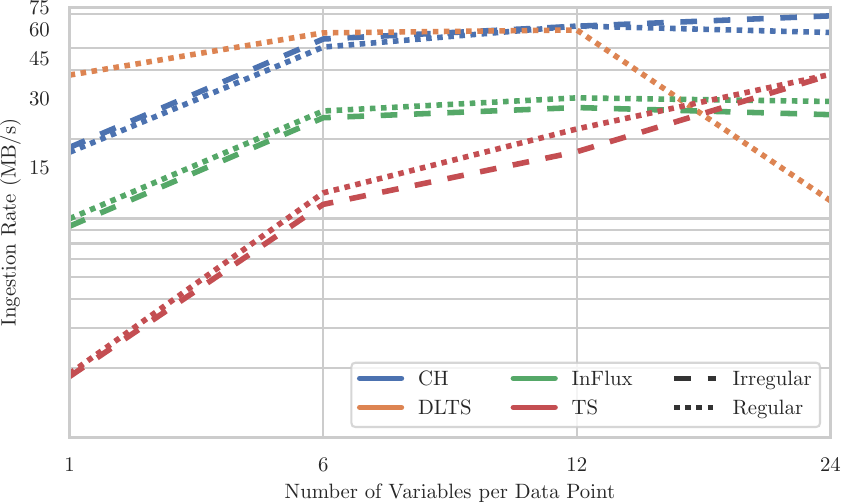}
    \caption{Ingestion Rate as function of Number of Variables per Data Point in a Time-Series at Batch Size 15000 and 64 Parallel Clients}
    \label{fig:lenght-multicard-b}
    \Description{}
\end{figure}

To understand the target TSDBs' performance under multi-variate time-series, we study peak ingestion rate (maximum possible over all batch sizes and parallel clients) as function of number of variables per data point, as show in \autoref{fig:lenght-multicard-b}.
To be fair with all TSDBs, we fix the batch size to 15000 and parallel clients to 64 where all TSDBs reach their ingestion rate upper bound, as discussed in section \ref{sec:monovariate}.
TS, CH, and IF show an incrementally increasing rate as we increase the number of variables per data point.
Despite TS's low ingestion rate of $\sim$\SI[per-mode=symbol]{1.89}{\mega\byte\per\second} for mono-variate regular time-series, its rate reaches \SI[per-mode=symbol]{38.21}{\mega\byte\per\second} for regular time-series and \SI[per-mode=symbol]{37.99}{\mega\byte\per\second} for irregular time-series at 24 variables exceeding the ingestion rate of IF.
IF's ingestion rate increase from \SI[per-mode=symbol]{8.98}{\mega\byte\per\second} at 1 variable per data point to \SI[per-mode=symbol]{26.47}{\mega\byte\per\second} at 6 variables for regular time-series, and from \SI[per-mode=symbol]{8.33}{\mega\byte\per\second} at 1 variable to \SI[per-mode=symbol]{24.73}{\mega\byte\per\second} at 6 variables for irregular time-series.
After 6 variables per data point, IF's ingestion rate slows to only \SI[per-mode=symbol]{30.2}{\mega\byte\per\second} at 12 variables and to \SI[per-mode=symbol]{29.12}{\mega\byte\per\second} at 24 variables for regular time-series, to \SI[per-mode=symbol]{27.39}{\mega\byte\per\second} at 12 variables and to \SI[per-mode=symbol]{25.51}{\mega\byte\per\second} at 24 variables for irregular time-series.
The ingestion rate of CH incrementally increases as we increase the number of variables per data point, until it achieves the highest ingestion rate at 24 variables for regular (up to \SI[per-mode=symbol]{68.76}{\mega\byte\per\second}) and irregular time-series (up to \SI[per-mode=symbol]{58.18}{\mega\byte\per\second}).
DLTS's ingestion rate incrementally increases from \SI[per-mode=symbol]{37.96}{\mega\byte\per\second} at 1 variable per data point to \SI[per-mode=symbol]{59.67}{\mega\byte\per\second} at 12 variables per data point.
However, at 24 variables, it significantly drops to \SI[per-mode=symbol]{10.75}{\mega\byte\per\second}.

Time-series regularity leads to a slight increase in ingestion rate for both IF~(up to $\sim$\SI[per-mode=symbol]{3}{\mega\byte\per\second}) and TS~(up to $\sim$\SI[per-mode=symbol]{5}{\mega\byte\per\second}).
Contrarily, the ingestion rate of CH drops up to $\sim$\SI[per-mode=symbol]{10}{\mega\byte\per\second} for regular time-series in comparison to irregular time-series.
This is a result of how every TSDB stores the ingested data on persistent disk.
For example, IF and TS are designed to group ingested data into shards in IF or chunks in TS which are both designed to store time-ordered data.
Such a storage engine architecture is optimized for regular time-series where writing the data to the disk leaves no empty slots in the shard/chunk.
% In the case of regular time-series, IF or TS fill the shards or the chunks contiguously and perform delta-encoding data compression like Gorilla for floats~\cite{gorilla}.
However, in the case of irregular time-series, data points are arriving at unpredictable intervals, and allocating a space for one data point leads to block fragmentation in the time-orderd shard/chunk, and, consequently lower ingestion rate.
On the other hand, CH manages ingested data as parts before merging the parts into the final columnar format.
The parts are constructed based on CH's sparse indexing algorithm where each part has a length equal to the sparse indexing algorithm (i.e. every how many data points to create a new data point).
This means 2 consecutive data points of an irregular time-series can end in different parts on the disk, and thus reducing disk contention on the part files through file parallelism.
In a regular time-series, it is most likely the 2 consecutive data points will end in the same part file, increasing disk contention and lowering ingestion rate.

\begin{figure}[t]
    \centering
    \includegraphics[width=1\linewidth]{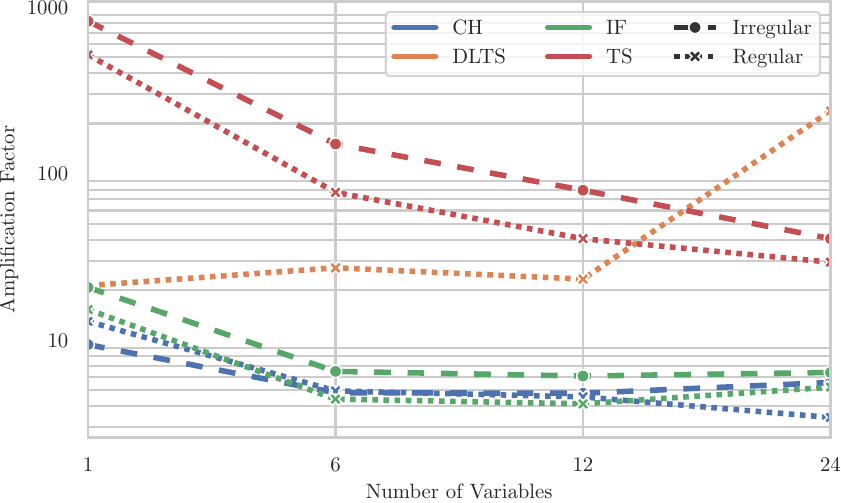}
    \caption{Disk Write Amplification Factor as function of Number of Variables per Data Point}
    \label{fig:bs5-15-multicard-DISK-CPU-efficiency-a}
    \Description{}
\end{figure}

To understand the performance patterns in TSDBs and how they are affected by disk write amplification, we measure their disk write.
\autoref{fig:bs5-15-multicard-DISK-CPU-efficiency-a} shows the disk write amplification factor as function of number of variables per data point.
The disk write amplification factor represents how much the ingested data has been amplified by the target TSDB.
As we increase the number of variables per data point, TS disk amplification factor gets reduced from $824\times$ at 1 variable per data point to $40.65\times$ at 24 variables for irregular time-series, and from $517.26\times$ at 1 variable to $29.35\times$ at 24 variables for regular time-series.
This is a result of diminishing disk write amplifications: while the amount of actually written data is increasing due to increasing the number of variables per data point, the size of the B-Tree index is constant, consequently, reducing disk write amplifications.
CH and IF maintains the lowest disk write amplification even at high number of variables per data point.
For example, IF's amplifications factors decrease from $20.71\times$ at 1 variable per data point to $6.38\times$ at 24 variables for irregular time-series, and from $15.24\times$ at 1 variable to $5.20\times$ at 24 variables for regular time-series.
Similarly, CH's amplifications factors decrease from $9.38\times$ at 1 variable per data point to $5.55\times$ at 24 variables for irregular time-series, and from $12.93\times$ at 1 variable to $3.42\times$ at 24 variables for regular time-series.
The LSM-based storage engines of CH and IF are optimizied for low disk usage and efficient resource usage.
The drop in amplifications factor is driven by storing the variables of one data point next to each other in TSDBs that group data on the disk by their relevant index.
This leads to fatter write operations (single write operation writing more data to disk) and consequently to efficient disk usage and higher ingestion rate.

Contrarily, DLTS's amplifications factors increase from $21.22\times$ at 1 variable to $238.12\times$ at 24 variables.
While increasing number of variables per data point, the ingestion rate sinks and disk write amplifications grows exponentially.
DLTS is a vector-based system designed for mono-variate time-series.
While DLTS can ingest multi-variate time-series, several vector files are created each representing a single variable.
Consequently, distinct variables are written to multiple vectors and saved in distinct blocks increasing random disk writes and thus lowering disk write throughput.
Thus, ingestion rate drops down due to decreasing disk efficiency and random disk writes.

\subsection{Dedicated Query Workloads}
We study the latency of the proposed queries in section \ref{sec:queries} for regular mono-variate and multi-variate time-series.
The TSDBs are populated with time-series data at first.
The queries are only executed after data compression has fully completed and without any other concurrent workload.
No parallel clients are employed in this section, a single client is querying the TSDB.

\subsubsection{Mono-Variate Time-Series}
\begin{figure}[ht!]
    \centering
    \includegraphics[width=1\linewidth]{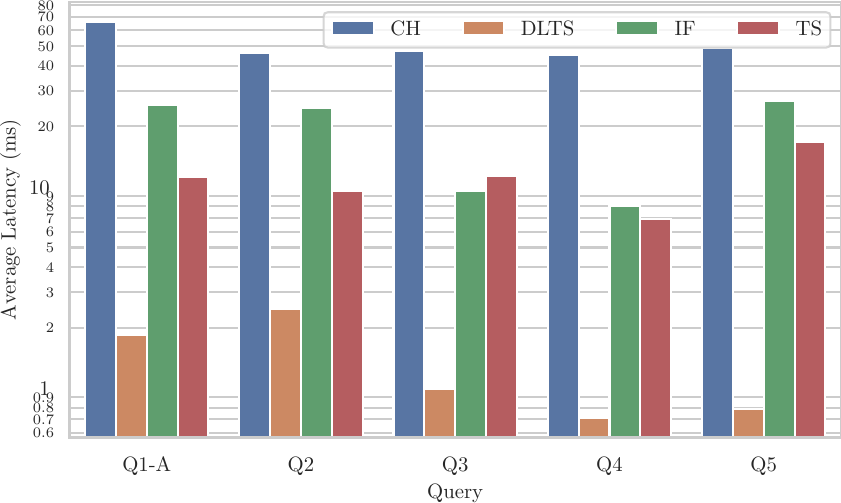}
    \caption{Query Latency for Mono-Variate Time-Series}
    \label{fig:query-card-reg}
    \Description{}
\end{figure}

Queries for mono-variate time-series include supported queries except Q1-B which is designed for multi-variate time-series.
\autoref{fig:query-card-reg} shows the average query latencies, measured in milliseconds on log scale, for the 4 TSDBs evaluated across the queries Q1-A through Q5.

Across all five queries, CH consistently exhibits the highest average latency, ranging from approximately \SI{45}{\milli\second} to \SI{48}{\milli\second} in Q2 to Q5 and hitting the maximum value of \SI{66}{\milli\second} to fetch raw time-series in Q1-A.
These query latencies make CH the slowest of the four systems irrespective of the query type.
To achieve high throughput, CH writes data to immutable sorted granules (sized at 8192 by default) for each column.
The column's data is then indexed sparsely, i.e. each entry in the sparse index points to the start of the column's granule but not to every data point inside the granule.
To answer a query, CH first searches the sparse index to find the corresponding granule and then performs a full-scan on the granule to find the target data point.
This architecture is optimized for bulk data operations over wide column sets, where the query result has coarse granularity, i.e. the result can be found quickly by finding the corresponding granules only without performing full-scans.
When query result becomes finely granuled, CH performs full-scans on the corresponding granules to find the target data points.
Consequently, this leads to higher query latency.

At the opposite end, DLTS delivers the lowest latency in every query class, and its advantage widens considerably in the later queries: latency falls from roughly \SI{2.47}{\milli\second} in Q2 to a minimum below \SI{1}{\milli\second} in Q3, Q4, and Q5.
DLTS's query latencies are notable.
Instead of searching for a data point using full-scans or indexing, DLTS exploits the regularity principle of time-series and saves the time series on the disk as a consecutively equidistant vector.
To find a range of data points, DLTS calculates the indices of the range extremities and extracts the corresponding vector slice.
Such an architecture is quite fast to find the corresponding data points, leading to very low query latencies.

TS and IF occupy the intermediate performance band, with TS generally outperforming IF in the first two query classes (\SI{11.15}{\milli\second} versus \SI{25.63}{\milli\second} in Q1-A, and \SI{9.55}{\milli\second} versus \SI{24.54}{\milli\second} in Q2).
The 2 TSDBs approximately converge in Q3 (\SI{11.3}{\milli\second} in TS vs. \SI{11.3}{\milli\second} in IF) and Q4 (\SI{6.95}{\milli\second} in TS vs. \SI{8.06}{\milli\second} in IF) before diverging again in Q5 ((\SI{16.7}{\milli\second} in TS vs. \SI{26.61}{\milli\second} in IF).
IF's TSM tree indexes data by a set of indexed tag (i.e. \emph{sensor\_id}).
This tag-based indexing model is highly efficient for queries that filter on moderate-cardinality tags, but query latency increase as cardinality grows~\cite{scits}.
On the other hand, TS efficiently exploits PostgreSQL efficient B-Tree indexing of composite primary key, i.e. \emph{(sensor\_id, time)}, as an index to achieve lower query latencies.

\subsubsection{Multi-Variate Time Series}
We study the latency as function of number of variables per data point to evaluate the query performance for multi-variate time-series.

\begin{figure}[!ht]
    \centering
    \includegraphics[width=1\linewidth]{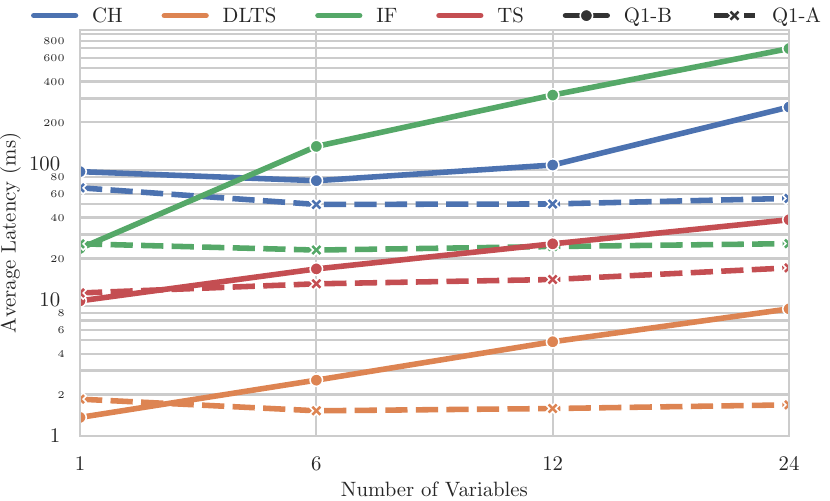}
    \caption{Query Latency for Raw Data Fetching a Single Variable (Q1-A) and All Variables (Q1-B) as function of Number of Variables per Data Point}
    \label{fig:querycard-agg-oneAll-a}
    \Description{}
\end{figure}

\autoref{fig:querycard-agg-oneAll-a} shows the query latency of four TSDBs while varying the number of variables per data point.
The evaluation examines the query performance of querying for a single variable vs querying a vector of variables per data point.
Q1-A is performed on multi-variate time-series by fetching the first variable of data point.
This specifically allows us to test the hypothesis if the TSDB needs to process all variables of the data point in case only one variable was queried for.

As seen in \autoref{fig:querycard-agg-oneAll-a}, none of the TSDBs show an increase in latency of Q1-A.
This shows that the number of variables per data point does not impact the performance of fetching a single variable of the data point.

As we increase the number of variables per data point, all TSDBs exhibit an increasing latency for Q1-B.
This is expected as the TSDBs have to read and process more data from the disk.
However, the reaction to this variation differ among TSDBs.
For example, IF shows an exponential increase in latency from \SI{23.69}{\milli\second} for 1 variable per data point to \SI{698.3}{\milli\second} at 24 variables per data point.
TS also shows a similar but softer increase in Q1-B latency.
TS ranges from \SI{9.76}{\milli\second} for 1 variable per data point to \SI{38.52}{\milli\second} at 24 variables, and CH ranges from \SI{87}{\milli\second} for 1 variable per data point to \SI{259.43}{\milli\second} at 24 variables.
Notably, DLTS's design efficiently scales from \SI{1.35}{\milli\second} for 1 variable per data point to only \SI{8.5}{\milli\second} at 24 variables.

\subsection{Mixed Workloads}
\begin{figure}[!ht]
    \centering
    \begin{subfigure}[t]{\linewidth}
        \centering
        \includegraphics[width=\linewidth]{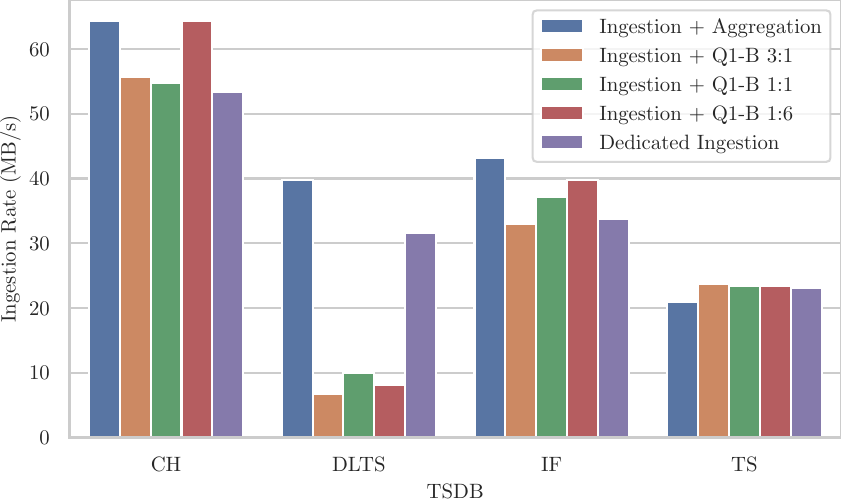}
        \caption{Ingestion Rate (15k-Sized Batches)}
        \label{fig:mixed-Ing-agg-RQperc-a}
    \end{subfigure}
    \begin{subfigure}[t]{\linewidth}
        \centering
        \includegraphics[width= \linewidth]{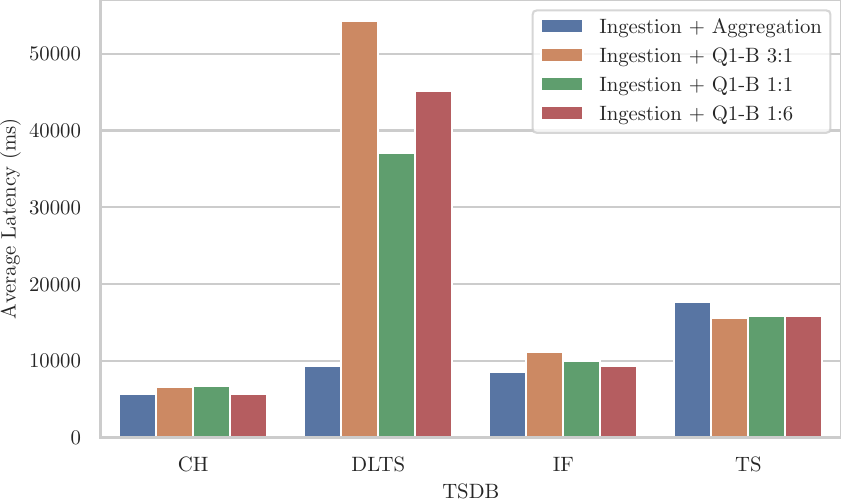}
        \caption{Queries Latencies}
        \label{fig:mixed-Ing-agg-RQperc-c}
    \end{subfigure}
    \caption{Performance of Mixed Workloads with Regular Time-Series and 256 Parallel Clients (128 Ingestion + 128 Querying)}
    \label{fig:mixed-Ing-agg-RQperc}
    \Description{}
\end{figure}

Realistic workloads are mixed involving both data ingestion and querying at the same time.
This section studies how the targeted TSDBs react to different mixed workloads.
\autoref{fig:mixed-Ing-agg-RQperc} shows the performance of the TSDBs while executing 256 parallel clients: 128 clients executing regular time-series ingestion workloads with batch size of 15k and another 128 clients querying the TSDB.
In addition to data ingestion, the mixed workloads are formed of 2 query workloads categories:
\begin{enumerate*}
    \item Aggregation Queries: a mix of Q3, Q4, and Q5 aggregation queries;
    \item Fractional Q1-B: a Q1-B query that is executed every N-inserted points to study how reading from disk impacts concurrent writing to disk, e.g. a \emph{Q1-B 3:1} query means that for each 3 inserted data points, the TSDB is queried for 1 data point, and \emph{Q1-B 1:6} for each inserted data point, the TSDB is queried for 6 data points.
\end{enumerate*}

\autoref{fig:mixed-Ing-agg-RQperc-a} shows the ingestion rate of the mixed workloads for each target TSDB in comparison to the dedicated ingestion workload.
While dedicated ingestion achieve an ingestion rate of \SI[per-mode=symbol]{53.37}{\mega\byte\per\second}, mixed workloads increase the ingestion rate from \SI[per-mode=symbol]{1.35}{\mega\byte\per\second} for mixed workload \emph{Ingestion + Q1-B 1:1} (i.e. insert a point for each queried point) and up to \SI[per-mode=symbol]{11}{\mega\byte\per\second} for mixed workload \emph{Ingestion + Aggregation Queries}.
IF and DLTS show the same pattern of increasing ingestion rate for mixed workloads by up to \SI[per-mode=symbol]{9.3}{\mega\byte\per\second} and \SI[per-mode=symbol]{8.27}{\mega\byte\per\second} respectively.

Diverging from previous results in section \ref{sec:dedicated-ing}, DLTS's ingestion rate drops lower than that of CH.
Another interesting observation is the sharp drop in ingestion rate of fractional Q1-B mixed workloads from \SI[per-mode=symbol]{31.51}{\mega\byte\per\second} in dedicated ingestion workloads to below \SI[per-mode=symbol]{10}{\mega\byte\per\second}.
We attribute both drops in ingestion rate (in fractional Q1-B mixed workloads and in comparison to CH) to the limitations of array-based storage in DLTS where ingestion and queries compete over the array.

On the other hand, TS is mostly neutral towards mixed workloads showing no significant increase or decrease in ingestion rate.
As discussed in section \ref{sec:monovariate}, TS is limited by PostgreSQL's concurrency model where an increasing number of clients harm the TSDB's performance.

\autoref{fig:mixed-Ing-agg-RQperc-c} shows the query latency in the mixed workloads for each target TSDB.
Extensive concurrency and mixed workloads increase the query latency by at least 2 order of magnitudes.
While achieving the highest ingestion rate, CH also achieves the lowest query latency in all mixed workloads ranging between \SI{5.72}{\second} for \emph{Ingestion + Q1-B 1:6} and \SI{6.73}{\second} for \emph{Ingestion + Q1-B 1:1}.
This shows the advantages of its MergeTree OLAP design that is specifically designed for such online analytics workloads.

IF shows a decreasing query latency in mixed workloads:
\SI{11.156}{\second} for \emph{Ingestion + Q1-B 3:1}, \SI{9.93}{\second} for \emph{Ingestion + Q1-B 1:1}, \SI{9.25}{\second} for \emph{Ingestion + Q1-B 1:6}, and \SI{8.55}{\second} for \emph{Ingestion + Aggregation}.
We notice that this highly correlates with the number of inserted data points while comparing the Q1-B workloads.
As we decrease the number of inserted data points for each queried data point, we notice that query latency is decreasing.
This shows that IF's query performance is impacted by the ingestion workload it has to handle.

Similar to ingestion in mixed workload, changing mixed workloads does not show an impact on query latency in TS.
The query latency of fractional Q1-B mixed workloads ranges between \SI{15.54}{\second} and \SI{15.57}{\second}.
The increased complexity of the aggregation queries increase the query latency to \SI{17.67}{\second}.

\begin{figure}[!ht]
    \centering
    \begin{subfigure}[t]{\linewidth}
        \centering
        \includegraphics[width=\linewidth]{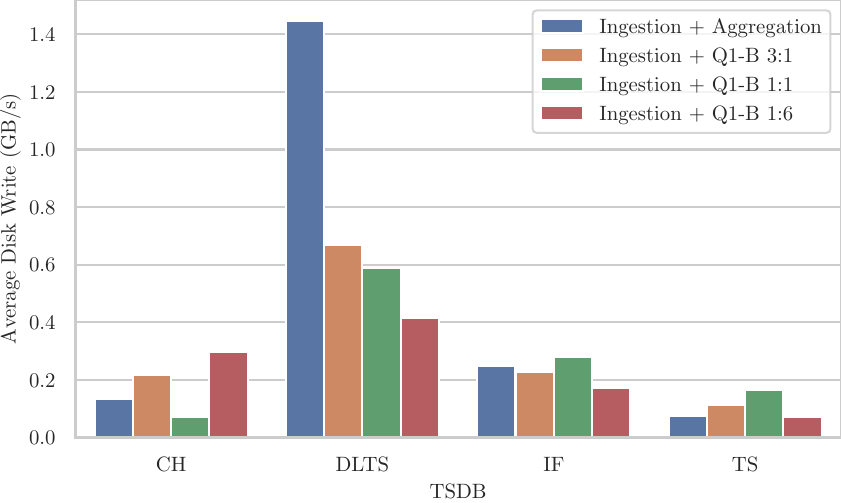}
        \caption{Disk Write (GB/s)}
        \label{fig:mixed-disk-wr}
    \end{subfigure}
    \begin{subfigure}[t]{\linewidth}
        \centering
        \includegraphics[width= \linewidth]{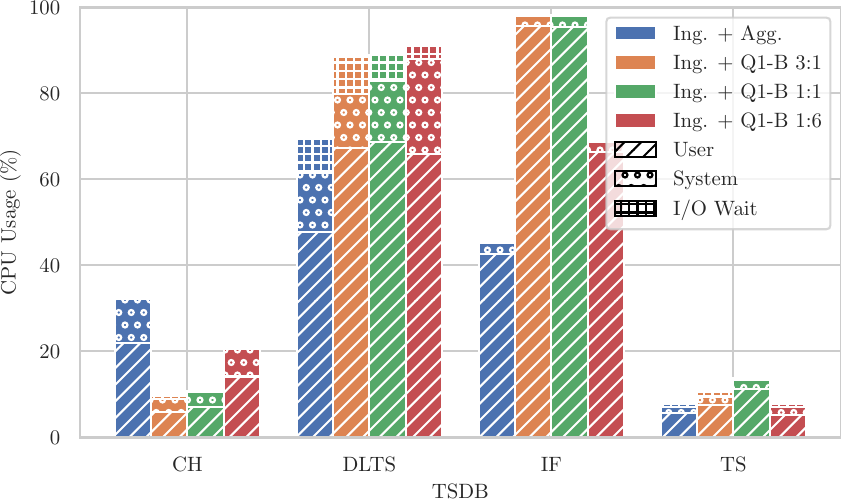}
        \caption{Splitted CPU Usage}
        \label{fig:mixed-cpu}
    \end{subfigure}
    \caption{System Metrics for Both Mixed Workloads}
    \label{fig:mixed-sysmetrics}
    \Description{}
\end{figure}

Further insights can be inferred from collected system metrics in SciTSv2.
\autoref{fig:mixed-sysmetrics} shows the collected system metrics for each TSDB with mixed workloads: \autoref{fig:mixed-disk-wr} shows Disk Write throughput in GB/s and \autoref{fig:mixed-cpu} shows the CPU usage splitted between time spent by the TSDB in the user-space, the system in kernel-space, and while waiting for I/O to finish.

CH, IF, and TS have the lowest Disk Write throughput ranging between \SI[per-mode=symbol]{0.07}{\giga\byte\per\second} and \SI[per-mode=symbol]{0.29}{\giga\byte\per\second}, \SI[per-mode=symbol]{0.17}{\giga\byte\per\second} and \SI[per-mode=symbol]{0.28}{\giga\byte\per\second}, and \SI[per-mode=symbol]{0.07}{\giga\byte\per\second} and \SI[per-mode=symbol]{0.16}{\giga\byte\per\second} respectively.
Considering CH's and IF's high ingestion rates ranging between \SI[per-mode=symbol]{53.37}{\mega\byte\per\second} and \SI[per-mode=symbol]{64}{\mega\byte\per\second} for CH and between \SI[per-mode=symbol]{53.37}{\mega\byte\per\second} and \SI[per-mode=symbol]{64}{\mega\byte\per\second} for IF as show in \autoref{fig:mixed-Ing-agg-RQperc-a}, they have the lowest Disk Write Amplification factor showing the most efficient disk usage.

\autoref{fig:mixed-cpu} shows that DLTS and IF have the highest CPU usage hitting 90\% and 98\% respectively.
While most of IF's CPU usage is spent in user-space (42.6\% to 95\%), 12.5\% to 22\% of DLTS's are spent in system time.
As we increase the number of read data points in fractional Q1-B workloads (increasing the disk read intensity of the query), we notice that the time intensity of CPU time spent in system increase: 12.54\% for \emph{Ingestion + Q1-B 3:1}, 14.25\% for \emph{Ingestion + Q1-B 1:1}, and 22\% for \emph{Ingestion + Q1-B 1:6}.
We also notice higher CPU usage while waiting for I/O operations to complete in DLTS reaching up to 8.53\%.
Other TSDBs do not demonstrate such behaviors.
Considering DLTS's higher query latency and lower ingestion rate, this shows that DLTS has a major bottleneck in mixed workloads.
This is likely to happen considering DLTS's design to re-write the value array while ingesting new data.

Among all TSDBs, CH shows the highest scalability.
Its OLAP design empowers CH's higher ingestion rate and lower query latency compared to other TSDBs while having a considerably low CPU usage and efficient disk usage.

\section{Conclusion}
\label{sec:conclusion}

The increasing volumes of time-series data in addition to the need to perform insightful data analysis pushed TSDBs as a critical component in the design of monitoring systems in big data use-cases like scientific infrastructure and IoT.
An insightful TSDB benchmark enables engineers to find bottlenecks in modern TSDBs.

This paper introduces SciTSv2, a novel benchmark for TSDBs with 6 dimensions: Connection Parallelism, Batch Data Ingestion, Regularity, Multi-Variate Time-Series, Mixed Workloads, and System Metrics.
We exploit the 6 dimensions of SciTSv2 to evaluate the performance of 4 TSDBs deeply and provide insights how they react on different workloads.
By systematically varying these dimensions independently, we exposed architectural limitations of the 4 targeted TSDBs.

\paragraph{Connection Parallelism \& Batch Ingestion}
Varying client count and batch size revealed granularity sensitivity.
DLTS excels at high concurrency and large batches but becomes unusable at small batches (\SI[per-mode=symbol, parse-numbers=false]{< 1}{\mega\byte\per\second}) due to full-array checkpoint rewrites.
CH, IF and TS saturate early; TS's PostgreSQL backend stalls beyond moderate parallelism, while IF spends excessive CPU sorting TSM files.

\paragraph{Time-Series Regularity}
Time-series regularity (equidistance) uncovered a counter-intuitive limitation: CH performs worse on regular data because its sparse-indexed parts concentrate writes into fewer files, increasing contention.
DLTS thrives on regularity but collapses on irregular data, exposing its narrow specialization.
IF and TS show only marginal gains, proving their sharding designs do not fundamentally exploit equidistance.
Without this dimension, one would falsely assume regularity is always beneficial.

\paragraph{Multi-Variate Series}
Sweeping the number of variables per data point exposed ingestion limits.
TS and CH scale positively with cardinality (amortizing index overhead), but TS still suffers extreme disk amplification.
IF peaks at 6 variables and stabilizes, revealing TSM's difficulty with wide schemas.
DLTS crashes at 24 variables because each additional variable is stored as a separate vector file, inducing random writes.
For queries, InfluxDB's latency grows exponentially, exposing tag-based indexing as unsuitable for high-dimensional lookups.

\paragraph{Mixed Ingestion-Query Workloads}
Concurrent read-write composition (varying ingestion/query ratios) revealed read-write contention as the dominant limitation.
DLTS's single-array architecture degrades severely—CPU system time spikes to 22\% where queries and ingestion checkpoints compete for the same vector file.
IF shows query latency directly proportional to insert intensity, indicating shared resource contention invisible in dedicated tests.
TS remains neutral but at low throughput, offering stability without performance.
CH alone maintains high ingestion and low latency under mixed loads, validating its OLAP design for real-time monitoring.

\paragraph{System Metrics}
Collecting CPU, memory, and disk I/O as first-class citizens transformed raw performance numbers into diagnostic evidence.
The root bottlenecks became clear: IF is CPU-bound (up to 98\% usage) from index sorting; TS is I/O-bound with disk amplification reaching 824$\times$ from per-chunk B-tree writes; DLTS is disk-bandwidth-bound (\SI[per-mode=symbol, parse-numbers=false]{2-2.5}{\giga\byte\per\second}) due to array rewriting; CH is highly resource-efficient.

The principal contribution of this framework lies in its capability to show different performance aspectes in TSDBs and in its capacity to systematically isolate and identify the root causes of architectural bottlenecks.
Each of the six dimensions serves as a controllable experimental variable, enabling controlled attribution of performance behavior to specific system characteristics.
By providing this diagnostic lens, SciTSv2 enables system architects to characterize storage engine behavior across diverse workloads, identify the specific conditions under which each architecture excels or degrades, and make data-driven selections that align with their operational requirements and scalability constraints.
\begin{acks}
    This work is supported by the Helmholtz Association and by the Ministry for Education and Research BMBF (grant numbers 05A23PMA, 05A23PX2, 05A23VK2 and 05A23WO6).
\end{acks}

\balance
\bibliographystyle{ACM-Reference-Format}
\bibliography{refs}

%%\appendix

\end{document}